# Relativistic symmetries in the Hulthén scalar-vector-tensor interactions

Ali Akbar Rajabi, Majid Hamzavi[*]

*Physics Department, Shahrood University of Technology, Shahrood, Iran*

[*]*Corresponding author: Tel.:+98 273 3395270, fax: +98 273 3395270*

Email: majid.hamzavi@gmail.com

**Abstract**

In the presence of spin and pseudospin (p-spin) symmetries, the approximate analytical bound states of the Dirac equation for scalar-vector-tensor Hulthén potentials are obtained with any arbitrary spin-orbit coupling number $\kappa$ using the Pekeris approximation. The Hulthén tensor interaction is studied instead of the commonly used Coulomb or linear terms. The generalized parametric Nikiforov-Uvarov (NU) method is used to obtain energy eigenvalues and corresponding wave functions in their closed forms. We show that tensor interaction removes degeneracies between spin and p-spin doublets. Some numerical results are also given.

# 1. Introduction

The spin and p-spin symmetry observed originally almost 40 years ago as a mechanism to explain different aspects of the nuclear structure is one of the most interesting phenomena in the relativistic quantum mechanics. Within the framework of Dirac equation, p-spin symmetry used to feature deformed nuclei, superdeformation and to establish an effective shell-model [1-4]. However, spin symmetry is relevant for mesons [5]. The spin symmetry occurs when the difference of the scalar $S(r)$ and vector $V(r)$ potentials is constant, i.e., $\Delta(r) = C_s$ and the p-spin symmetry occurs when the sum of the scalar and vector potentials is constant, i.e., $\Sigma(r) = C_{ps}$ [6-7]. The p-spin symmetry refers to a quasi-degeneracy of single nucleon doublets with non-relativistic quantum number $(n, l, j = l + 1/2)$ and $(n-1, l+2, j = l + 3/2)$, where $n$, $l$ and $j$ are single nucleon radial, orbital and total angular quantum numbers, respectively [8-9]. The total angular momentum $j = \tilde{l} + \tilde{s}$ with $\tilde{l} = l+1$ is a pseudo-angular momentum and $\tilde{s}$ is p-spin angular momentum [10-14]. Tensor potentials were introduced into the Dirac equation with the substitution $\vec{p} \to \vec{p} - im\omega\beta.\hat{r}U(r)$ and a spin-orbit coupling is added to the Dirac Hamiltonian [15-16]. For more review on tensor interaction, one can see Refs. [17-25] that authors used different potential and different kind of tensor potential like Coulomb and linear ones, but here we study a tensor potential in the Hulthén form.

The Hulthén potential has the form

$$v(r) = -v_0 \frac{e^{-\delta r}}{1 - e^{-\delta r}} \qquad (1)$$

where $\delta$ is the screening parameter which is used for determining the range of the Hulthén potential. The parameter $v_0$ represents $\delta Z e^2$, where $Ze$ is the charge of the nucleon [26]. The intensity of the Hulthén potential is denoted by $v_0$ under the condition of $\delta > 0$. This potential has been used in several branches of physics and its discrete and continuum states have been studied by a variety of techniques such as the algebraic perturbation calculations which are based upon the dynamical group structure SO(2,1) [27], the formalism of supersymmetric quantum mechanics within the framework of the variational method [28], the supersymmetry and shape invariance property [29], the asymptotic iteration method [30,31], the NU method [32] and the approach proposed by Biedenharn for the Dirac-Coulomb problem [33,34].

Ikhdair and sever obtained the energy eigenvalues and wave functions of the Dirac particle in the field of the scalar and vector Hulthén potentials including a Coulomb-like tensor interaction [10]. The advantage of this work is to obtain the energy eigenvalues and wave functions of the Dirac particle in the fields of the scalar and vector Hulthén potentials including a Hulthén tensor interaction in view of spin and p-spin symmetries.

The structure of this paper is organized as follows. In Section 2, we briefly introduce the Dirac equation with scalar, vector and tensor potentials with arbitrary spin-orbit coupling number $\kappa$ including tensor interaction under spin and p-spin symmetric limits. The generalized parametric NU method is presented in Section 3. The energy eigenvalue equations and corresponding wave functions in terms of the Jacobi polynomials are obtained in Section 4. We also give some remarks and numerical results. We end with conclusion in Section 5.

## 2. Dirac Equation including Tensor Coupling

The Dirac equation for fermionic massive spin-$1/2$ particles moving in an attractive scalar $S(r)$, repulsive vector $V(r)$ and tensor $U(r)$ potentials is (in units $\hbar = c = 1$)

$$[\vec{\alpha}.\vec{p} + \beta(M + S(r)) - i\beta\vec{\alpha}.\hat{r}U(r)]\psi(\vec{r}) = [E - V(r)]\psi(\vec{r}), \qquad (2)$$

where $E$ is the relativistic energy of the system, $\vec{p} = -i\vec{\nabla}$ is the three-dimensional momentum operator and $M$ is the mass of the fermionic particle. Further, $\vec{\alpha}$ and $\beta$ are the $4 \times 4$ Dirac matrices given by

$$\vec{\alpha} = \begin{pmatrix} 0 & \vec{\sigma} \\ \vec{\sigma} & 0 \end{pmatrix}, \qquad \beta = \begin{pmatrix} I & 0 \\ 0 & -I \end{pmatrix}, \qquad (3)$$

where $I$ is $2 \times 2$ unitary matrix and $\vec{\sigma}$ are three-vector spin matrices

$$\sigma_1 = \begin{pmatrix} 0 & 1 \\ 1 & 0 \end{pmatrix}, \quad \sigma_2 = \begin{pmatrix} 0 & -i \\ i & 0 \end{pmatrix}, \quad \sigma_3 = \begin{pmatrix} 1 & 0 \\ 0 & -1 \end{pmatrix}. \qquad (4)$$

The total angular momentum operator $\vec{J}$ and spin-orbit $K = (\vec{\sigma}.\vec{L} + 1)$, where $\vec{L}$ is orbital angular momentum of the spherical nucleons commute with the Dirac

Hamiltonian. The eigenvalues of spin-orbit coupling operator are $\kappa = \left(j+\frac{1}{2}\right) > 0$ and $\kappa = -\left(j+\frac{1}{2}\right) < 0$ for unaligned spin $j = l - \frac{1}{2}$ and the aligned spin $j = l + \frac{1}{2}$, respectively. $(H^2, K, J^2, J_z)$ can be taken as the complete set of the conservative quantities. Thus, the spinor wave functions can be classified according to their angular momentum $j$, spin-orbit quantum number $\kappa$ and the radial quantum number $n$ can be written as follows

$$\psi_{n\kappa}(\vec{r}) = \begin{pmatrix} f_{n\kappa}(\vec{r}) \\ g_{n\kappa}(\vec{r}) \end{pmatrix} = \begin{pmatrix} \frac{F_{n\kappa}(r)}{r} Y_{jm}^{l}(\theta,\varphi) \\ i \frac{G_{n\kappa}(r)}{r} Y_{jm}^{\tilde{l}}(\theta,\varphi) \end{pmatrix}, \tag{5}$$

where $f_{n\kappa}(\vec{r})$ is the upper (large) component and $g_{n\kappa}(\vec{r})$ is the lower (small) component of the Dirac spinors. $Y_{jm}^{l}(\theta,\varphi)$ and $Y_{jm}^{\tilde{l}}(\theta,\varphi)$ are spin and p-spin spherical harmonics, respectively, and $m$ is the projection of the angular momentum on the $z$-axis. Substituting Eq. (5) into Eq. (2) and using the following relations

$$(\vec{\sigma}.\vec{A})(\vec{\sigma}.\vec{B}) = \vec{A}.\vec{B} + i\vec{\sigma}.(\vec{A}\times\vec{B}), \tag{7a}$$

$$(\vec{\sigma}.\vec{P}) = \vec{\sigma}.\hat{r}\left(\hat{r}.\vec{P} + i\frac{\vec{\sigma}.\vec{L}}{r}\right), \tag{7b}$$

together with the following properties

$$\begin{aligned}
(\vec{\sigma}.\vec{L})Y_{jm}^{\tilde{l}}(\theta,\phi) &= (\kappa-1)Y_{jm}^{\tilde{l}}(\theta,\phi), \\
(\vec{\sigma}.\vec{L})Y_{jm}^{l}(\theta,\phi) &= -(\kappa-1)Y_{jm}^{l}(\theta,\phi), \\
(\vec{\sigma}.\hat{r})Y_{jm}^{\tilde{l}}(\theta,\phi) &= -Y_{jm}^{l}(\theta,\phi), \\
(\vec{\sigma}.\hat{r})Y_{jm}^{l}(\theta,\phi) &= -Y_{jm}^{\tilde{l}}(\theta,\phi),
\end{aligned} \tag{8}$$

one obtains two coupled differential equations for upper and lower radial wave functions $F_{n\kappa}(r)$ and $G_{n\kappa}(r)$ as

$$\left(\frac{d}{dr} + \frac{\kappa}{r} - U(r)\right)F_{n\kappa}(r) = (M + E_{n\kappa} - \Delta(r))G_{n\kappa}(r), \tag{9a}$$

$$\left(\frac{d}{dr} - \frac{\kappa}{r} + U(r)\right)G_{n\kappa}(r) = (M - E_{n\kappa} + \Sigma(r))F_{n\kappa}(r). \tag{9b}$$

where

$$\Delta(r) = V(r) - S(r), \tag{10a}$$

$$\Sigma(r) = V(r) + S(r), \tag{10b}$$

are the difference and the sum potentials, respectively. Eliminating $F_{n\kappa}(r)$ and $G_{n\kappa}(r)$ from Eqs. (9), we finally obtain the following two Schrödinger-like differential equations for the upper and lower radial spinor components, respectively:

$$
\begin{aligned}
&\left[\frac{d^2}{dr^2} - \frac{\kappa(\kappa+1)}{r^2} + \frac{2\kappa}{r}U(r) - \frac{dU(r)}{dr} - U^2(r)\right]F_{n\kappa}(r) \\
&+ \frac{\frac{d\Delta(r)}{dr}}{M + E_{n\kappa} - \Delta(r)}\left(\frac{d}{dr} + \frac{\kappa}{r} - U(r)\right)F_{n\kappa}(r) \\
&= \left[(M + E_{n\kappa} - \Delta(r))(M - E_{n\kappa} + \Sigma(r))\right]F_{n\kappa}(r),
\end{aligned}
\tag{11}
$$

$$
\begin{aligned}
&\left[\frac{d^2}{dr^2} - \frac{\kappa(\kappa-1)}{r^2} + \frac{2\kappa}{r}U(r) + \frac{dU(r)}{dr} - U^2(r)\right]G_{n\kappa}(r) \\
&+ \frac{\frac{d\Sigma(r)}{dr}}{M - E_{n\kappa} + \Sigma(r)}\left(\frac{d}{dr} - \frac{\kappa}{r} + U(r)\right)G_{n\kappa}(r) \\
&= \left[(M + E_{n\kappa} - \Delta(r))(M - E_{n\kappa} + \Sigma(r))\right]G_{n\kappa}(r)
\end{aligned}
\tag{12}
$$

where $\kappa(\kappa-1) = \tilde{l}(\tilde{l}+1)$ and $\kappa(\kappa+1) = l(l+1)$. The quantum number $\kappa$ is related to the quantum numbers for spin symmetry $l$ and p-spin symmetry $\tilde{l}$ as

$$\kappa = \begin{cases} -(l+1) = -(j+\frac{1}{2}) & (s_{1/2}, p_{3/2}, etc.) \quad j = l + \frac{1}{2}, \text{ aligned spin}(\kappa < 0) \\ +l = +(j+\frac{1}{2}) & (p_{1/2}, d_{3/2}, etc.) \quad j = l - \frac{1}{2}, \text{ unaligned spin}(\kappa > 0), \end{cases}$$

and the quasidegenerate doublet structure can be expressed in terms of a p-spin angular momentum $\tilde{s} = 1/2$ and pseudo-orbital angular momentum $\tilde{l}$, which can be defined as

$$\kappa = \begin{cases} -\tilde{l} = -(j+\frac{1}{2}) & (s_{1/2}, p_{3/2}, etc.) \quad j = \tilde{l} - \frac{1}{2}, \text{ alinged p-spin}(\kappa < 0) \\ +(\tilde{l}+1) = +(j+\frac{1}{2}) & (d_{3/2}, f_{5/2}, etc.) \quad j = \tilde{l} + \frac{1}{2}, \text{ unaligned p-spin}(\kappa > 0), \end{cases}$$

where $\kappa = \pm 1, \pm 2, \ldots$. For example, $(1s_{1/2}, 0d_{3/2})$ and $(1p_{3/2}, 0f_{5/2})$ can be considered as p-spin doublets.

## 2.1. P-spin Symmetric Limit

Ginocchio showed that there is a connection between p-spin symmetry and the time component of a vector potential and the scalar potential are nearly equal, i.e., $S(r) \approx -V(r)$ [7]. After that, Meng et al. derived that if $\frac{d[V(r)+S(r)]}{dr} = \frac{d\Sigma(r)}{dr} = 0$ or $\Sigma(r) = C_{ps} =$ constant, p-spin symmetry is exact in the Dirac equation [35-38]. In this part, we take attractive scalar $S(r)$, repulsive vector $V(r)$ and tensor $U(r)$ potentials as;

$$V(r) = V_0 \frac{e^{-\delta r}}{1-e^{-\delta r}}, \quad S(r) = -S_0 \frac{e^{-\delta r}}{1-e^{-\delta r}} \tag{13}$$

$$U(r) = -U_0 \frac{e^{-\delta r}}{1-e^{-\delta r}} \tag{14}$$

and the difference of scalar and vector potentials becomes

$$\Delta(r) = \Delta_0 \frac{e^{-\delta r}}{1-e^{-\delta r}}, \tag{15}$$

where $\Delta_0 = V_0 + S_0$. Under this symmetry, from Eq. (12), one obtains

$$\left[ \frac{d^2}{dr^2} - \frac{\kappa(\kappa-1)}{r^2} - 2\kappa U_0 \frac{e^{-\delta r}}{r(1-e^{-\delta r})} + U_0 \delta \left( \frac{e^{-\delta r}}{1-e^{-\delta r}} + \frac{e^{-2\delta r}}{(1-e^{-\delta r})^2} \right) \right.$$
$$\left. - U_0^2 \frac{e^{-2\delta r}}{(1-e^{-\delta r})^2} - \tilde{\gamma}\Delta_0 \frac{e^{-\delta r}}{1-e^{-\delta r}} - \tilde{\beta}^2 \right] G_{n\kappa}(r) = 0 \tag{16}$$

where $\tilde{\gamma} = E_{n\kappa} - M - C_{ps}$, $\tilde{\beta}^2 = (M + E_{n\kappa})(M - E_{n\kappa} + C_{ps})$. Also, $\kappa = -\tilde{l}$ and $\kappa = \tilde{l} + 1$ for $\kappa < 0$ and $\kappa > 0$, respectively.

## 2.2. Spin Symmetric Limit

In the spin symmetry we require $\frac{d\Delta(r)}{dr} = 0$ or $\Delta(r) = C_s =$ constant [35-38] and $\Sigma(r)$ as

$$\Sigma(r) = \Sigma_0 \frac{e^{-\delta r}}{1-e^{-\delta r}}, \tag{17}$$

where $\Sigma_0 = -(S_0 - V_0)$. Then Eq. (11) becomes

$$\left[\frac{d^2}{dr^2}-\frac{\kappa(\kappa+1)}{r^2}-2\kappa U_0 \frac{e^{-\delta r}}{r(1-e^{-\delta r})}-U_0\delta\left(\frac{e^{-\delta r}}{1-e^{-\delta r}}+\frac{e^{-2\delta r}}{(1-e^{-\delta r})^2}\right)\right.$$
$$\left.-U_0^2\frac{e^{-2\delta r}}{(1-e^{-\delta r})^2}+\gamma\Sigma_0\frac{e^{-\delta r}}{1-e^{-\delta r}}-\beta^2\right]G_{n\kappa}(r)=0 \tag{18}$$

where $\gamma = M + E_{n\kappa} - C_s$ and $\beta^2 = (M - E_{n\kappa})(M + E_{n\kappa} - C_s)$. Also, $\kappa = l$ and $\kappa = -l - 1$ for $\kappa < 0$ and $\kappa > 0$, respectively.

Because of the spin-orbit coupling term, Eqs. (15) and (16) can not be solved analytically, we shall use the shifted approximation [32] in order to deal with the spin-orbit coupling term as

$$\frac{1}{r^2} \approx \delta^2 \frac{e^{-2\delta r}}{(1-e^{-\delta r})^2}, \tag{19a}$$

or equivalently

$$\frac{1}{r^2} \approx \delta \frac{e^{-\delta r}}{1-e^{-\delta r}}, \tag{19b}$$

The exponential numerator in Eq. (19) is expanded for small values of $r$ and higher order terms are ignored up to first-order term [32]. In Fig. 1, we plot the centrifugal term and its approximation with two different values of $\delta$, to show the accuracy of the approximation.

## 3. Parametric NU Method

This powerful mathematical tool is usually used to solve second order differential equations. Let us consider the following differential equation [39]

$$\psi_n''(s)+\frac{\tilde{\tau}(s)}{\sigma(s)}\psi_n'(s)+\frac{\tilde{\sigma}(s)}{\sigma^2(s)}\psi_n(s)=0, \tag{20}$$

where $\sigma(s)$ and $\tilde{\sigma}(s)$ are polynomials, at most of second degree, and $\tilde{\tau}(s)$ is a first-degree polynomial. To make the application of the NU method simpler and direct without need to check the validity of solution. We present a shortcut for the method. Hence, firstly we write the general form of the Schrödinger-like equation (20) in a more general form as [39,40]

$$\psi_n''(s)+\left(\frac{c_1-c_2 s}{s(1-c_3 s)}\right)\psi_n'(s)+\left(\frac{-p_2 s^2 + p_1 s - p_0}{s^2(1-c_3 s)^2}\right)\psi_n(s)=0, \tag{21}$$

satisfying the wave functions

$$\psi_n(s) = \phi(s) y_n(s). \tag{22}$$

Secondly, we compare (21) with its counterpart (20) to obtain the following parameter values,

$$\tilde{\tau}(s) = c_1 - c_2 s, \quad \sigma(s) = s(1 - c_3 s), \quad \tilde{\sigma}(s) = -p_2 s^2 + p_1 s - p_0, \tag{23}$$

Now, following the NU method [39], we obtain the energy equation [40]

$$c_2 n - (2n+1)c_5 + (2n+1)\left(\sqrt{c_9} - c_3\sqrt{c_8}\right) + n(n-1)c_3 + c_7 + 2c_3 c_8 - 2\sqrt{c_8 c_9} = 0, \tag{24}$$

and the corresponding wave functions

$$\rho(s) = s^{c_{10}} (1 - c_3 s)^{c_{11}}, \quad \phi(s) = s^{c_{12}} (1 - c_3 s)^{c_{13}}, \quad c_{12} > 0, \; c_{13} > 0,$$

$$y_n(s) = P_n^{(c_{10}, c_{11})} (1 - 2c_3 s), \quad c_{10} > -1, \; c_{11} > -1,$$

$$\psi_{n\kappa}(s) = N_{n\kappa} s^{c_{12}} (1 - c_3 s)^{c_{13}} P_n^{(c_{10}, c_{11})} (1 - 2c_3 s). \tag{25}$$

where $P_n^{(\mu,\nu)}(x)$, $\mu > -1$, $\nu > -1$, and $x \in [-1,1]$ are Jacobi polynomials with the following constants:

$$c_4 = \frac{1}{2}(1 - c_1), \qquad\qquad c_5 = \frac{1}{2}(c_2 - 2c_3),$$

$$c_6 = c_5^2 + p_2, \qquad\qquad c_7 = 2c_4 c_5 - p_1,$$

$$c_8 = c_4^2 + p_0, \qquad\qquad c_9 = c_3(c_7 + c_3 c_8) + c_6,$$

$$c_{10} = c_1 + 2c_4 - 2\sqrt{c_8} - 1 > -1, \qquad c_{11} = 1 - c_1 - 2c_4 + \frac{2}{c_3}\sqrt{c_9} > -1, \; c_3 \neq 0,$$

$$c_{12} = c_4 - \sqrt{c_8} > 0, \qquad\qquad c_{13} = -c_4 + \frac{1}{c_3}(\sqrt{c_9} - c_5) > 0, \; c_3 \neq 0, \tag{26}$$

where $c_{12} > 0$, $c_{13} > 0$ and $s \in [0, 1/c_3]$, $c_3 \neq 0$.

In the special case when $c_3 = 0$, the wave function (22) becomes

$$\lim_{c_3 \to 0} P_n^{(c_{10}, c_{11})} (1 - 2c_3 s) = L_n^{c_{10}}(c_{11} s), \quad \lim_{c_3 \to 0} (1 - c_3 s)^{c_{13}} = e^{c_{13} s},$$

$$\psi(s) = N s^{c_{12}} e^{c_{13} s} L_n^{c_{10}}(c_{11} s). \tag{27}$$

## 4. Relativistic Bound States for the Hulthén-Like and Tensor Potentials
## 4.1. P-spin Symmetric Solution

Substituting Eq. (19) into Eq. (16) gives

$$\left[\frac{d^2}{dr^2} - \kappa(\kappa-1)\delta^2 \frac{e^{-2\delta r}}{(1-e^{-\delta r})^2} - 2\kappa U_0 \delta \frac{e^{-2\delta r}}{(1-e^{-\delta r})^2} + U_0 \delta\left(\frac{e^{-\delta r}}{1-e^{-\delta r}} + \frac{e^{-2\delta r}}{(1-e^{-\delta r})^2}\right)\right.$$
$$\left. -U_0^2 \frac{e^{-2\delta r}}{(1-e^{-\delta r})^2} - \tilde{\gamma}\Delta_0 \frac{e^{-\delta r}}{1-e^{-\delta r}} - \tilde{\beta}^2\right] G_{n\kappa}(r) = 0 \quad (28)$$

and further making the change of variables $s = e^{-\delta r}$, one obtains

$$\frac{d^2 G_{n\kappa}(s)}{ds^2} + \frac{1-s}{s(1-s)}\frac{dG_{n\kappa}(s)}{ds}$$
$$+ \frac{1}{\delta^2 s^2(1-s)^2}\left[-\tilde{A}s^2 + \tilde{B}s(1-s) - \tilde{\beta}^2(1-s)^2\right] G_{n\kappa}(s) = 0 \quad (29)$$

where

$$\tilde{A} = \kappa(\kappa-1)\delta^2 + U_0(U_0 + 2\kappa\delta - \delta) \quad (30a)$$

$$\tilde{B} = -\tilde{\gamma}\Delta_0 + U_0 \delta \quad (30b)$$

Comparing Eq. (28) and Eq. (21), we can easily obtain the coefficients $c_i$ ($i = 1, 2, 3$) and analytical expressions $p_j$ ($j = 0, 1, 2$) as follows

$$c_1 = 1, \qquad p_0 = \frac{\tilde{\beta}^2}{\delta^2}$$

$$c_2 = 1, \qquad p_1 = \frac{\tilde{B} + 2\tilde{\beta}^2}{\delta^2}$$

$$c_3 = 1, \qquad p_2 = \frac{\tilde{A} + \tilde{B} + \tilde{\beta}^2}{\delta^2} \quad (30)$$

The values of coefficients $c_i$ ($i = 4, 5, ..., 13$) are found from Eq. (26) as below

$$c_4 = 0, \qquad c_5 = -\frac{1}{2},$$

$$c_6 = \frac{1}{4} + \frac{\tilde{A} + \tilde{B} + \tilde{\beta}^2}{\delta^2}, \qquad c_7 = -\frac{\tilde{B} + 2\tilde{\beta}^2}{\delta^2}$$

$$c_8 = \frac{\tilde{\beta}^2}{\delta^2}, \qquad c_9 = p_2 - p_1 + p_0 + \frac{1}{4} = \left(\kappa - \frac{1}{2}\right)^2 + \frac{U_0}{\delta}\left(\frac{U_0}{\delta} + 2\kappa - 1\right)$$

$$c_{10} = -2\sqrt{\frac{\tilde{\beta}^2}{\delta^2}}, \qquad c_{11} = 2\tilde{\eta}_\kappa,$$

$$c_{12} = -\sqrt{\frac{\tilde{\beta}^2}{\delta^2}}, \qquad c_{13} = \tilde{\eta}_\kappa + \frac{1}{2}, \qquad (31)$$

where

$$\tilde{\eta}_\kappa = \sqrt{\left(\kappa - \frac{1}{2}\right)^2 + \frac{U_0}{\delta}\left(\frac{U_0}{\delta} + 2\kappa - 1\right)} \qquad (32)$$

By using (24), we can obtain the energy eigenvalues of the Hulthén potential with p-spin symmetric limit as

$$\left(n + \frac{1}{2} + \sqrt{\left(\kappa - \frac{1}{2}\right)^2 + \frac{U_0}{\delta}\left(\frac{U_0}{\delta} + 2\kappa - 1\right)} - \sqrt{\frac{\tilde{\beta}^2}{\delta^2}}\right)^2$$

$$= \frac{-\tilde{\gamma}\Delta_0 + \tilde{\beta}^2}{\delta^2} + \left(\kappa - \frac{1}{2}\right)^2 + \frac{U_0}{\delta}\left(\frac{U_0}{\delta} + 2\kappa\right) - \frac{1}{4}. \qquad (33)$$

Some numerical results of above equation are given in table 1. In table 1, we obtain numerical results in the p-spin symmetry in the presence and absence of Tensor Hulthén form potential. We took a set of parameter values, $M = 5.0 \text{fm}^{-1}$, $C_{ps} = 0$, $V_0 = 2.5 \text{ fm}^{-1}$, $S_0 = 2.9 \text{fm}^{-1}$ and $\delta = 0.1 \text{fm}^{-1}$. It is observed the existence of the degeneracy in the following doublets $(1s_{1/2}, 0d_{3/2}), (1p_{3/2}, 0f_{5/2}), (1d_{5/2}, 0g_{7/2}), (1f_{7/2}, 0h_{9/2})$ and so on, in the absence of the tensor potential ($U_0 = 0$) and they are considered as p-spin doublets.

In Fig. 2, we have investigated the effect of the tensor potential on the p-spin doublet splitting by considering the following pairs of orbital: $(1d_{5/2}, 0g_{7/2}), (2f_{7/2}, 1h_{9/2})$. From Fig. 2, we observe that in the case of $U_0 = 0$ (no tensor interaction), members of p-spin doublets have same energy. However, in the presence of the tensor potential $U_0 \neq 0$, these degeneracies are removed.

To find the corresponding wave functions, referring to table 1 and relations in (23), we find the functions

$$\rho(s) = s^{-2\sqrt{\frac{\tilde{\beta}^2}{\delta^2}}}(1-s)^{2\tilde{\eta}_\kappa},$$

$$\phi(s) = s^{-\sqrt{\frac{\tilde{\beta}^2}{\delta^2}}}(1-s)^{\tilde{\eta}_\kappa + \frac{1}{2}},$$

$$y_n(s) = P_n^{(-2\sqrt{\frac{\tilde{\beta}^2}{\delta^2}}, 2\tilde{\eta}_\kappa)}(1-2s), \qquad (34)$$

By using $G_{n\kappa}(s) = \phi(s) y_n(s)$, we get the lower component of the Dirac spinor from relation (25) as

$$G_{n\kappa}(s) = \tilde{N}_{n\kappa} s^{-\sqrt{\frac{\tilde{\beta}^2}{\delta^2}}} (1-s)^{\tilde{\eta}_\kappa + \frac{1}{2}} P_n^{(-2\sqrt{\frac{\tilde{\beta}^2}{\delta^2}}, 2\tilde{\eta}_\kappa)}(1-2s), \qquad (35)$$

or equivalently

$$G_{n\kappa}(r) = \tilde{N}_{n\kappa} e^{-\sqrt{\tilde{\beta}^2} r} (1-e^{-\delta r})^{\tilde{\eta}_\kappa + \frac{1}{2}} P_n^{(-2\sqrt{\frac{\tilde{\beta}^2}{\delta^2}}, 2\tilde{\eta}_\kappa)}(1-2e^{-\delta r}), \qquad (36)$$

where $\tilde{N}_{n\kappa}$ is normalization constant. The upper component of the Dirac spinor from Eq. (9b) can be calculated as

$$F_{n\kappa}(r) = \frac{1}{M - E_{n\kappa} + C_{ps}} \left( \frac{d}{dr} - \frac{\kappa}{r} + U(r) \right) G_{n\kappa}(r) \qquad (37)$$

where $E \neq M + C_{ps}$.

**4.2. Spin Symmetric Solution**

In this subsection we will obtain the energy eigenvalues and the corresponding wave functions for the spin symmetric limit of the Hulthén potential, i.e. solutions of Eq. (18). Substituting Eq. (19) into Eq. (18) gives

$$\left[ \frac{d^2}{dr^2} - \kappa(\kappa+1)\delta^2 \frac{e^{-2\delta r}}{(1-e^{-\delta r})^2} - 2\kappa U_0 \delta \frac{e^{-2\delta r}}{(1-e^{-\delta r})^2} - U_0 \delta \left( \frac{e^{-\delta r}}{1-e^{-\delta r}} + \frac{e^{-2\delta r}}{(1-e^{-\delta r})^2} \right) \right.$$
$$\left. - U_0^2 \frac{e^{-2\delta r}}{(1-e^{-\delta r})^2} + \gamma \Sigma_0 \frac{e^{-\delta r}}{1-e^{-\delta r}} - \beta^2 \right] F_{n\kappa}(r) = 0 \qquad (38)$$

and further making the change of variables $s = e^{-\delta r}$, one obtains

$$\frac{d^2 G_{n\kappa}(s)}{ds^2} + \frac{1-s}{s(1-s)} \frac{dG_{n\kappa}(s)}{ds}$$
$$+ \frac{1}{\delta^2 s^2 (1-s)^2} \left[ -As^2 + Bs(1-s) - \tilde{\beta}^2 (1-s)^2 \right] F_{n\kappa}(s) = 0 \qquad (39)$$

where

$$A = \kappa(\kappa+1)\delta^2 + U_0(U_0 + 2\kappa\delta + \delta) \qquad (40a)$$

$$B = \gamma \Sigma_0 - U_0 \delta \qquad (40b)$$

On the other hand, to avoid repetition in our solution, the energy states of Eq. (11), in the spin symmetric case:

$$\left(n+\frac{1}{2}+\sqrt{\left(\kappa+\frac{1}{2}\right)^2+\frac{U_0}{\delta}\left(\frac{U_0}{\delta}+2\kappa+1\right)}-\sqrt{\frac{\beta^2}{\delta^2}}\right)^2$$

$$=\frac{-\gamma\Sigma_0+\beta^2}{\delta^2}+\left(\kappa+\frac{1}{2}\right)^2+\frac{U_0}{\delta}\left(\frac{U_0}{\delta}+2\kappa\right)-\frac{1}{4}. \tag{41}$$

Also, we get the upper component of the Dirac spinor as

$$F_{n\kappa}(r)=N_{n\kappa}e^{-\sqrt{\beta^2}\,r}\left(1-e^{-\delta r}\right)^{\eta_\kappa+\frac{1}{2}}P_n^{(-2\sqrt{\frac{\beta^2}{\delta^2}},\,2\eta_\kappa)}\left(1-2e^{-\delta r}\right), \tag{42}$$

where $N_{n\kappa}$ is normalization constant and

$$\eta_\kappa=\sqrt{\left(\kappa+\frac{1}{2}\right)^2+\frac{U_0}{\delta}\left(\frac{U_0}{\delta}+2\kappa+1\right)}. \tag{43}$$

The lower component of the Dirac spinor from Eq. (9a) can be calculated as

$$G_{n\kappa}(r)=\frac{1}{M+E_{n\kappa}-C_s}\left(\frac{d}{dr}+\frac{\kappa}{r}-U(r)\right)F_{n\kappa}(r), \tag{44}$$

where $E\neq -M+C_s$.

In table 2, we use same parameters as before subsection. We can observe that every pair of orbitals ($np_{1/2}$, $np_{3/2}$), ($nd_{3/2}$, $nd_{5/2}$) and ($nf_{5/2}$, $nf_{7/2}$) has the same energy in the absence of the tensor potential ($U_0=0$). Thus, they can be viewed as the spin doublets, i.e., the state $1p_{1/2}$ with $n=1$ and $\kappa=1$ forms a spin doublet with the $1p_{3/2}$ state with $n=1$ and $\kappa=-2$. On the other hand, in the presence of the tensor potential ($U_0\neq 0$), one can notice that degeneracy between every pair of spin doublets is removed.

In Fig. 3, we have investigated the effect of the tensor potential on the spin doublet splitting by considering the following pairs of orbital: $(1p_{3/2},1p_{1/2})$, $(1f_{7/2},1f_{5/2})$ and one can observe that the results obtained in the spin symmetric limit resemble the ones observed in the p-spin symmetric limit.

## 5. Conclusion

By using the Pekeris approximation, we have obtained bound state solutions of the Hulthén scalar-vector-tensor potentials under p-spin and spin symmetries with arbitrary spin-orbit coupling number $\kappa$. This work is an improved work of Ref. [10] that the authors used a different kind of tensor potential as Coulomb-like potential.

But, here we studied the problem with Hulthén tensor potential. We use a powerful mathematical tool, i.e. generalized parametric NU method, to solve second order differential equations. Also, from numerical results and figures, it is found that tensor interaction removes degeneracies between each pairs of p-spin or spin doublets.

**Table 1.** The bound state energy eigenvalues in unit of $fm^{-1}$ of the p-spin symmetry Hulthén potential for several values of $n$ and $\kappa$ with $U_0 = 0.1$.

| $\tilde{l}$ | $n, \kappa < 0$ | $(l, j)$ | $E_{n,\kappa<0}$ $U_0 \neq 0$ | $E_{n,\kappa<0}$ $U_0 = 0$ | $n-1, \kappa > 0$ | $(l+2, j+1)$ | $E_{n-1,\kappa>0}$ $U_0 \neq 0$ | $E_{n-1,\kappa>0}$ $U_0 = 0$ |
|---|---|---|---|---|---|---|---|---|
| 1 | 1, -1 | $1s_{1/2}$ | 4.999528842 | 4.998866406 | 0, 2 | $0d_{3/2}$ | 4.998896324 | 4.998866406 |
| 2 | 1, -2 | $1p_{3/2}$ | 4.999147492 | 4.998655904 | 0, 3 | $0f_{5/2}$ | 4.998720844 | 4.998655904 |
| 3 | 1, -3 | $1d_{5/2}$ | 4.998896324 | 4.998512210 | 0, 4 | $0g_{7/2}$ | 4.998591717 | 4.998512210 |
| 4 | 1, -4 | $1f_{7/2}$ | 4.998720844 | 4.998407889 | 0, 5 | $0h_{9/2}$ | 4.998492820 | 4.998407889 |
| 1 | 2, -1 | $2s_{1/2}$ | 4.998552452 | 4.997499195 | 1, 2 | $1d_{3/2}$ | 4.997268501 | 4.997499195 |
| 2 | 2, -2 | $2p_{3/2}$ | 4.997816234 | 4.996981851 | 1, 3 | $1f_{5/2}$ | 4.996848860 | 4.996981851 |
| 3 | 2, -3 | $2d_{5/2}$ | 4.997268501 | 4.996589219 | 1, 4 | $1g_{7/2}$ | 4.996517976 | 4.996589219 |
| 4 | 2, -4 | $2f_{7/2}$ | 4.996848860 | 4.996281439 | 1, 5 | $1h_{9/2}$ | 4.996250655 | 4.996281439 |

**Table 2.** The bound state energy eigenvalues in unit of $fm^{-1}$ of the spin symmetry Hulthén potential for several values of $n$ and $\kappa$ with $U_0 = 0.1$.

| $l$ | $n, \kappa < 0$ | $(l, j = l+1/2)$ | $E_{n,\kappa<0}$ $U_0 \neq 0$ | $E_{n,\kappa<0}$ $U_0 = 0$ | $n, \kappa > 0$ | $(l, j = l-1/2)$ | $E_{n,\kappa>0}$ $U_0 \neq 0$ | $E_{n,\kappa>0}$ $U_0 = 0$ |
|---|---|---|---|---|---|---|---|---|
| 1 | 0, -2 | $0p_{3/2}$ | −4.999038017 | −4.999752463 | 0, 1 | $0p_{1/2}$ | −4.999559444 | −4.999752463 |
| 2 | 0, -3 | $0d_{5/2}$ | −4.999445752 | −4.999751647 | 0, 2 | $0d_{3/2}$ | −4.999611783 | −4.999751647 |
| 3 | 0, -4 | $0f_{7/2}$ | −4.999559444 | −4.999751236 | 0, 3 | $0f_{5/2}$ | −4.999641705 | −4.999751236 |
| 4 | 0, -5 | $0g_{9/2}$ | −4.999611783 | −4.999750989 | 0, 4 | $0g_{7/2}$ | −4.999661028 | −4.999750989 |
| 1 | 1, -2 | $1p_{3/2}$ | −4.998475156 | −4.998659482 | 1, 1 | $1p_{1/2}$ | −4.998134589 | −4.998659482 |
| 2 | 1, -3 | $1d_{5/2}$ | −4.998252842 | −4.998456496 | 1, 2 | $1d_{3/2}$ | −4.998061278 | −4.998456496 |
| 3 | 1, -4 | $1f_{7/2}$ | −4.998134589 | −4.998327074 | 1, 3 | $1f_{5/2}$ | −4.998011397 | −4.998327074 |
| 4 | 1, -5 | $1g_{9/2}$ | −4.998061278 | −4.998237543 | 1, 4 | $1g_{7/2}$ | −4.997975269 | −4.998237543 |

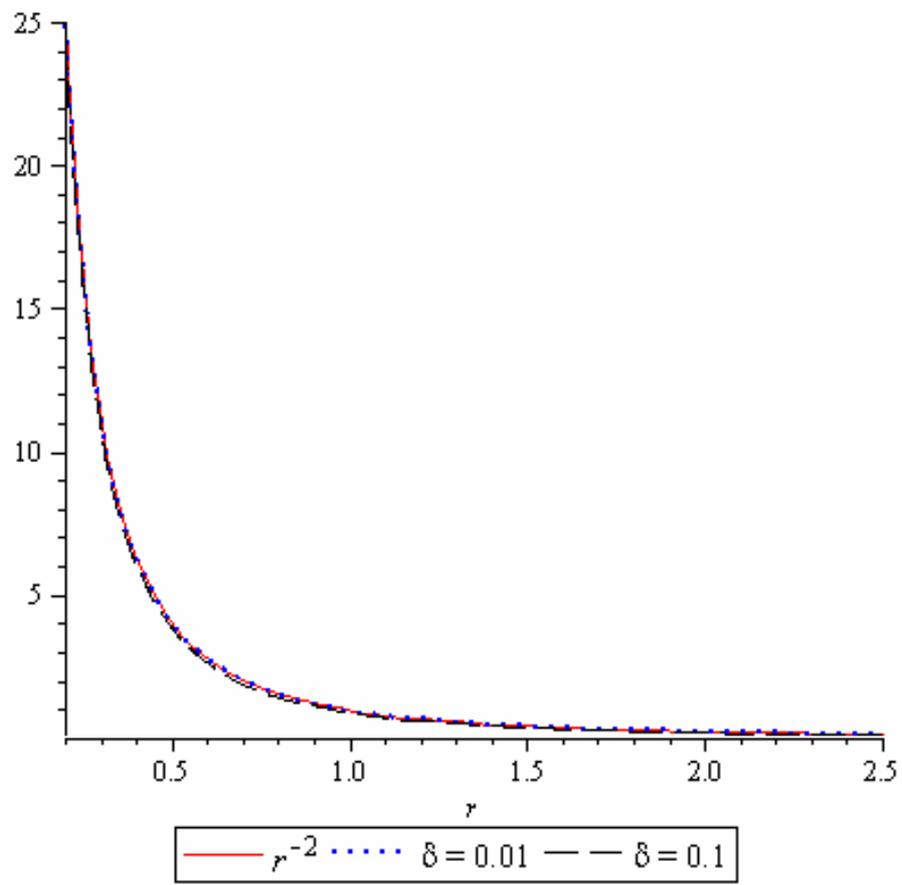

**Fig 1.** The centrifugal term $r^{-2}$ (red solid curve) and its approximation in Eq. (19a) with $\delta = 0.01 fm^{-1}$ (blue dot curve) and $\delta = 0.1 fm^{-1}$ (black dash curve).

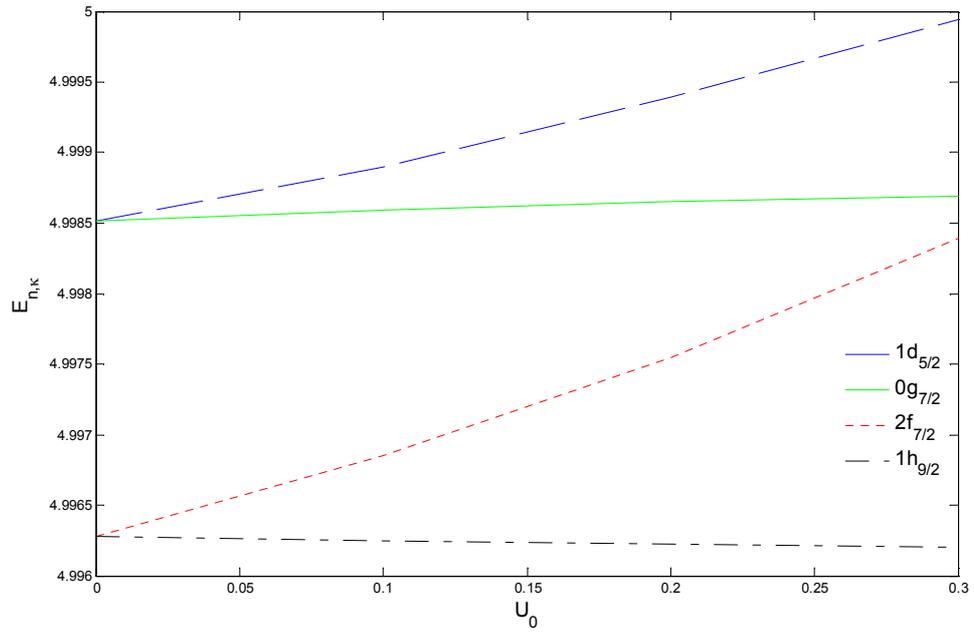

**Fig 2.** Contribution of the tensor potential parameter to the energy levels in the case of p-spin symmetry.

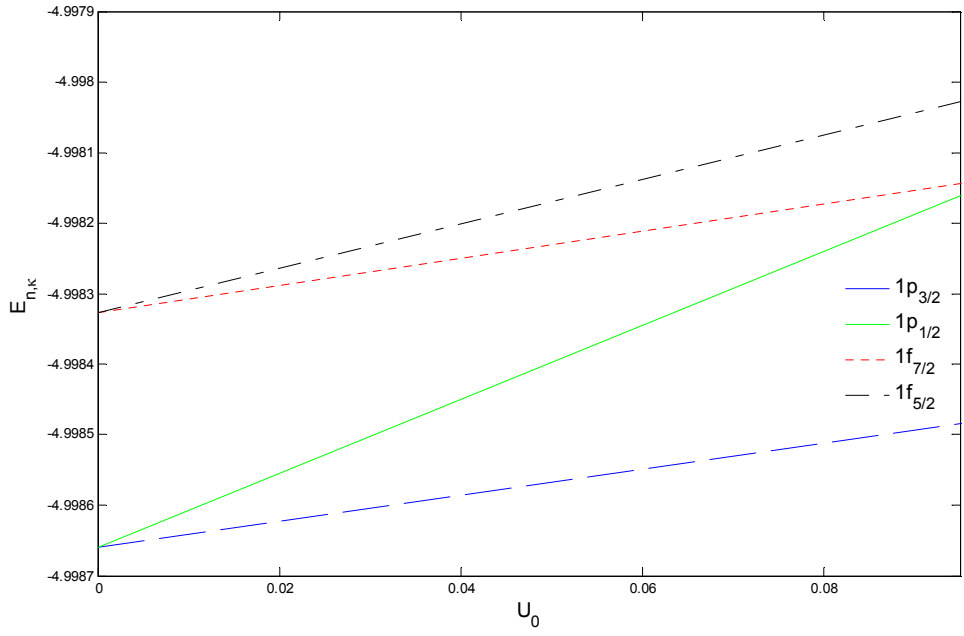

**Fig 3.** Contribution of the tensor potential parameter to the energy levels in the case of spin symmetry.